\documentclass[preprint,aps]{revtex4}
\usepackage{graphicx}
\usepackage{amsmath,mathrsfs}
\begin{document}
\bibliographystyle{revtex}
\title{Effect of Phase Shift in Shape Changing Collision of Solitons in 
Coupled Nonlinear Schr{\"o}dinger Equations}
%\subtitle{Do you have a subtitle?\\ If so, write it here}
\author{T. Kanna and M. Lakshmanan}
\affiliation{Centre for Nonlinear Dynamics, Department of Physics, Bharathidasan
University, Tiruchirapalli 620 024, India}
%\date{}
\newcommand{\be}{\begin{equation}}
\newcommand{\ee}{\end{equation}}
\newcommand{\al}{\alpha}
\newcommand{\bear}{\begin{eqnarray}}
\newcommand{\eear}{\end{eqnarray}}
\newcommand{\sech}{\mbox{ sech}}
\begin{abstract}
Soliton interactions in systems modelled by 
coupled nonlinear Schr{\"o}dinger (CNLS) 
equations and encountered in phenomena such as wave propagation in optical fibers 
and photorefractive media possess unusual features : shape changing intensity 
redistributions, amplitude dependent phase shifts and relative separation 
distances. We demonstrate these properties in the case of integrable 2-CNLS 
equations. As a simple example, we consider the stationary two-soliton solution 
which is equivalent to the so-called partially coherent soliton (PCS) solution 
discussed much in the recent literature. 
\end{abstract}
     \maketitle
%\authorrunning{Kanna, Lakshmanan}
%\titlerunning{Effect of Phase Shift in Shape Changing Collision of Solitons 
%in CNLS Equations}
%
\section{Introduction}
%\label{intro}
The study on the formation of optical solitons and their intriguing 
interaction properties is becoming one of the frontier areas of 
research in nonlinear dynamics due to their potential 
technological applications[1,2]. 
Indeed optical solitons are 
becoming desirable candidates in long distance optical communication systems,
in optical devices and in optical computers.
In a mathematical sense these solitons appear basically
 as solutions of integrable coupled 
nonlinear Schr{\"o}dinger (CNLS) type equations. For example, 
the intense electromagnetic 
wave propagation in a birefringent fiber is governed by 
the following set of 2-CNLS equations for the envelopes $q_1$ and $q_2$,
 which is in general nonintegrable,
\begin{eqnarray}
iq_{1z}+q_{1tt}+2\mu(|q_1|^2+B|q_2|^2)q_1  =  0,\nonumber\\
iq_{2z}+q_{2tt}+2\mu(|q_2|^2+B|q_1|^2)q_2  =  0,
\end{eqnarray} 
where z and t represent the normalized distance along the fiber 
and the retarded time respectively, $\mu$ represents the strength of 
nonlinearity and 
$B=\frac{2+2sin^2 \theta}{2+cos^2 \theta}$ is the cross phase modulation 
coupling parameter ($\theta$: ellipticity angle).
However this system becomes integrable for $B=1$. The resulting set of
equations
\begin{eqnarray}
iq_{1z}+q_{1tt}+2\mu(|q_1|^2+|q_2|^2)q_1  &=& 0,\nonumber\\ 
iq_{2z}+q_{2tt}+2\mu(|q_1|^2+|q_2|^2)q_2  &=&  0, 
\end{eqnarray}
is the celebrated Manakov equation[3]. In a 
recent work Radhakrishnan, Lakshmanan and Hietarinta [4] have revealed the 
fact that the soliton solutions of the integrable 2-CNLS (Manakov) 
equations undergo a fascinating 
shape-changing collision, resulting in a redistribution 
of intensity between the two solitons in the two modes, which is not observed in
the scalar nonlinear Schr{\"o}dinger (NLS) equation which exhibits only pure
elastic collision without any redistribution of intensities of solitons. 
Consequently, Jakubowski, Steiglitz and Squier[5]
have pointed out the possibility of using this 
phenomenon 
in constructing logic gates and in a very recent work[6] Steiglitz constructed 
such gates including the universal NAND gate, thereby showing the theoretical 
possibility of constructing all optical computers without interconnecting
discrete components in a homogeneous bulk nonlinear optical medium. Also Yang[7]
has studied the effect of additional perturbations on these solitons using
perturbation theory. Further such integrable CNLS 
equations arise in the context of spatial solitons as well which are 
receiving renewed attention for their formation at very low optical powers 
in photorefractive medium[8]. The present authors have extended the results 
of 2-CNLS system to 3- and N-CNLS equations [9]. 

\noindent All the above investigations mostly concentrate on the effect of changes 
in the amplitude (polarization) and the consequent effect on the 
energy redistribution between the modes of the solitons. 
So far not much attention has
been paid to the role of phases during optical soliton interaction. In 
this report, we point out the significance of 
amplitude dependent phase shift / relative seperation distance
 involved in the interaction process (sec.2) 
responsible for the shape change of solitons during collision along with the 
changes in the amplitudes (polarization) of the modes in the Manakov system. It
may be noted that such amplitude dependent phase shifts do not occur in the case
of scalar NLS equation. As a simple 
example, we consider (sec.3) the role of phase shifts / relative separation
 distances for stationary 2-soliton case and point out that this solution is nothing 
but the so-called stationary partially coherent soliton (PCS) in the recent 
literature[10]. 
\section{Soliton Interaction in 2-CNLS System}
To start with let us consider briefly the nature of one- and two-soliton 
solutions [4,11]. Multisoliton solutions of CNLS equations and 
their interactions 
will be considered elsewhere[12].
 %\label{sec:1}
%and \cite{RefJ}
\subsection{One-soliton solution}
The one-soliton solution to eq.(2) can be given in terms of 
three arbitrary 
complex parameters $\alpha_1^{(1)}$,$\alpha_1^{(2)}$ and $k_1$ as[4,11]
\begin{eqnarray}
\left(
\begin{array}{c}
q_1\\
q_2 
\end{array}
\right)
& = &
\left(
\begin{array}{c}
\alpha_1^{(1)}\\
\alpha_1^{(2)}
\end{array}
\right)\frac{e^{\eta_1}}{1+e^{\eta_1+\eta_1^*+R}},\\
 &=& 
\left(
\begin{array}{c}
A_1 \\
A_2
\end{array}
\right)
\frac{k_{1R}e^{i\eta_{1I}}}{\mbox{cosh}\,(\eta_{1R}+\frac{R}{2})}, 
\end{eqnarray}  
where $\eta_i=k_i(t+ik_iz)$, $i=1$, where $k_i=k_{iR}+ik_{iI}$, $k_{iR}$ and
$k_{iI}$ represent the real and imaginary parts of $k_i$. 
Here $\sqrt{\mu} (A_1,A_2)=\frac{(\alpha_1^{(1)},\alpha_1^{(2)})} 
{\sqrt{|\alpha_1^{(1)}|^2+|\alpha_1^{(2)}|^2}}$                                                                                                	
%Here $\sqrt{\mu} (A_1,A_2)=\frac{\sqrt{\mu}(\alpha_1^{(1)},\alpha_1^{(2)})} 
%{(\mu(|\alpha_1^{(1)}|^2+|\alpha_1^{(2)}|^2))^{(1/2)}}$ 
represents the unit polarization vector,
$k_{1R}A_j$,$\;$j=1,2 gives the amplitude of the $j$th mode and 
$2k_{1I}$ is the soliton  velocity. 
\subsection{Two-soliton solution}
Introducing six complex parameters $\alpha_1^{(1)}$, $\alpha_2^{(1)}$, 
$\alpha_1^{(2)}$, $\alpha_2^{(2)}$, $k_1$ and $k_2$, the two-soliton 
solution can be given as [4,11]
\begin{eqnarray}
q_1&=&\frac{\alpha_1^{(1)}e^{\eta_1}+\alpha_2^{(1)}e^{\eta_2}
+e^{\eta_1+\eta_1^*+\eta_2+\delta_1}+e^{\eta_1+\eta_2+\eta_2^*
+\delta_2}}
{D},\nonumber\\
q_2 &= & \frac{\alpha_1^{(2)}e^{\eta_1}+\alpha_2^{(2)}e^{\eta_2}
+e^{\eta_1+\eta_1^*+\eta_2+\delta_1'}+e^{\eta_1+\eta_2+\eta_2^*
+\delta_2'}}
{D},\nonumber
\end{eqnarray}
where
\begin{eqnarray}
D &=&  1+e^{\eta_1+\eta_1^*+R_1}
+e^{\eta_1+\eta_2^*+\delta_0}
 +e^{\eta_1^*+\eta_2+\delta_0^*}+e^{\eta_2+\eta_2^*+R_2}\nonumber\\
&&+e^{\eta_1+\eta_1^*+\eta_2+\eta_2^*+R_3}.
\end{eqnarray}
Here 
\begin{eqnarray}
\eta_i&=&k_i(t+ik_iz),\;\;
e^{\delta_0} = \frac{\kappa_{12}}{k_1+k_2^*},\;\;
e^{R_j} = \frac{\kappa_{jj}}{k_j+k_j^*},
\nonumber\\
e^{\delta_1}&=&\frac{k_1-k_2}{(k_1+k_1^*)(k_1^*+k_2)}
(\alpha_1^{(1)}\kappa_{21}-\alpha_2^{(1)}\kappa_{11}),\nonumber\\
e^{\delta_2}&=&\frac{k_2-k_1}{(k_2+k_2^*)(k_1+k_2^*)}
(\alpha_2^{(1)}\kappa_{12}-\alpha_1^{(1)}\kappa_{22}),\nonumber\\
e^{\delta_1^{'}}&=& \frac{k_1-k_2}{(k_1+k_1^*)(k_1^*+k_2)}
(\alpha_1^{(2)}\kappa_{21}-\alpha_2^{(2)}\kappa_{11}),\nonumber\\
e^{\delta_2^{'}}&=& \frac{k_2-k_1}{(k_2+k_2^*)(k_1+k_2^*)}
(\alpha_2^{(2)}\kappa_{12}-\alpha_1^{(2)}\kappa_{22}),\nonumber\\
e^{R_3}&=&  \frac{|k_1-k_2|^2}{(k_1+k_1^*)(k_2+k_2^*)|k_1+k_2^*|^2}
 (\kappa_{11}\kappa_{22}-\kappa_{12}\kappa_{21})\nonumber\\
\mbox{and}\;
%\begin{eqnarray}
\kappa_{ij}& = &\frac{\mu(\alpha_i^{(1)}\alpha_j^{(1)*}+
\alpha_i^{(2)}\alpha_j^{(2)*})}{k_i+k_j^*},\;i,j=1,2.\nonumber
\end{eqnarray}
The above two-soliton solution represents the interaction of two coupled 
one solitons. The scenario behind this interaction is that there is an 
intensity redistribution among the two modes of the two solitons along 
with an amplitude dependent phase shift and relative separation 
distance[4,9,11]. In order to understand the nature of the collisions we can 
consider the following cases for $k_{1I}$$>$$k_{2I}$: 

\noindent (a) $k_{1R}>0$, $k_{2R}>0$ 
(b) $k_{1R}>0$, $k_{2R}<0$
(c) $k_{1R}<0, k_{2R}>0$ 
(d) $k_{1R}<0, k_{2R}<0$. Similarly, one can consider four cases for 
$ k_{1I}<k_{2I} $. In all these cases an asymptotic analysis 
$(z\rightarrow \pm\infty)$ reveals the following structures. 

\noindent 1) \underline{Limit{\bf $z \rightarrow -\infty$}}

\noindent (a) $Soliton1$ :
\begin{eqnarray}
\left(
\begin{array}{c}
q_1\\
q_2 
\end{array}
\right)
\rightarrow
\left(
\begin{array}{c}
A_1^{1-} \\
A_2^{1-}
\end{array}
\right)
k_{1R}e^{i\eta_{1I}}{\mbox{sech}\,\left(\eta_{1R}+\hat{\phi}^{1-}\right)}, 
\end{eqnarray}
(b) $Soliton2$:
\begin{eqnarray}
\left(
\begin{array}{c}
q_1\\
q_2 
\end{array}
\right)
\rightarrow
\left(
\begin{array}{c}
A_1^{2-} \\
A_2^{2-}
\end{array}
\right)
k_{2R}e^{i\eta_{2I}}{\mbox{sech}\,\left(\eta_{2R}+\hat{\phi}^{2-}\right)}. 
\end{eqnarray}
2) \underline{ Limit{\bf $z \rightarrow +\infty$}}

\noindent (a) $Soliton1$ :
\begin{eqnarray}
\left(
\begin{array}{c}
q_1\\
q_2 
\end{array}
\right)
\rightarrow
\left(
\begin{array}{c}
A_1^{1+} \\
A_2^{1+}
\end{array}
\right)
k_{1R}e^{i\eta_{1I}}{\mbox{sech}\,\left(\eta_{1R}+\hat{\phi}^{1+}\right)}, 
\end{eqnarray}
(b) $Soliton2$:
\begin{eqnarray}
\left(
\begin{array}{c}
q_1\\
q_2 
\end{array}
\right)
\rightarrow
\left(
\begin{array}{c}
A_1^{2+} \\
A_2^{2+}
\end{array}
\right)
k_{2R}e^{i\eta_{2I}}{\mbox{sech}\,\left(\eta_{2R}+\hat{\phi}^{2+}\right)}. 
\end{eqnarray}
Here the various quantities corresponding to the four cases for 
$k_{1I}$$>$$k_{2I}$ are given below.\\
%%%%%%%%%%%%%%%%
\underline{Case(a)$k_{1R}>0$,$k_{2R}>0$}:
\begin{eqnarray}
\left(
\begin{array}{c}
A_1^{1-}\\
A_2^{1-} 
\end{array}
\right)
&=&
\left(
\begin{array}{c}
\alpha_1^{(1)} \\
\alpha_1^{(2)}
\end{array}
\right)e^{-R_1/2},\\
\left(
\begin{array}{c}
A_1^{1+}\\
A_2^{1+} 
\end{array}
\right)
&=&
\left(
\begin{array}{c}
e^{\delta_2} \\
e^{\delta_2'}
\end{array}
\right)e^{-(R_2+R_3)/2},\\
\left(
\begin{array}{c}
A_1^{2-}\\
A_2^{2-} 
\end{array}
\right)
&=&
\left(
\begin{array}{c}
e^{\delta_1} \\
e^{\delta_1'}
\end{array}
\right)e^{-(R_1+R_3)/2},\\
\left(
\begin{array}{c}
A_1^{2+}\\
A_2^{2+} 
\end{array}
\right)
&=&
\left(
\begin{array}{c}
\alpha_2^{(1)} \\
\alpha_2^{(2)}
\end{array}
\right)e^{-R_2/2},
\end{eqnarray}
$\hat{\phi}^{1-}=\frac{R_1}{2}$, $\hat{\phi}^{1+}
=\frac{R_3-R_2}{2}$,$\hat{\phi}^{2-}=\frac{R_3-R_1}{2}$ and 
$\hat{\phi}^{2+}=\frac{R_2}{2}$.
%%%%%%%%%%%%%%%
\underline{Case(b)$k_{1R}>0$,$k_{2R}<0$}:
\begin{eqnarray}
\left(
\begin{array}{c}
A_1^{j-}\\
A_2^{j-} 
\end{array}
\right)
&=&
\left(
\begin{array}{c}
e^{\delta_{l}} \\
e^{\delta_{l}'}
\end{array}
\right)e^{-(R_{l}+R_3)/2},\\
\left(
\begin{array}{c}
A_1^{j+}\\
A_2^{j+} 
\end{array}
\right)
&=&
\left(
\begin{array}{c}
\alpha_j^{(1)} \\
\alpha_j^{(2)}
\end{array}
\right)e^{-R_j/2},
\end{eqnarray}
$\hat{\phi}^{j-}=\frac{R_3-R_{l}}{2}$ and $\hat{\phi}^{j+}
=\frac{R_j}{2}$, $j=1,2$.\\
%%%%%%%%%%%%%%%
%%%%%%%%%%%%%%
\underline{Case(c) $k_{1R}<0$,$k_{2R}>0$}:
\begin{eqnarray}
\left(
\begin{array}{c}
A_1^{j-}\\
A_2^{j-} 
\end{array}
\right)
&=&
\left(
\begin{array}{c}
\alpha_j^{(1)} \\
\alpha_j^{(2)}
\end{array}
\right)e^{-R_j/2},\\
\left(
\begin{array}{c}
A_1^{j+}\\
A_2^{j+} 
\end{array}
\right)
&=&
\left(
\begin{array}{c}
e^{\delta_{l}} \\
e^{\delta_{l}'}
\end{array}
\right)e^{-(R_{l}+R_3)/2},
\end{eqnarray}
$\hat{\phi}^{j-}=\frac{R_j}{2}$ and $\hat{\phi}^{j+}
=\frac{R_3-R_{l}}{2}$.\\
In both the cases (b) and (c) $l=j+(-1)^{j+1}$, $j=1,2$.
%%%%%%%%%%%%%%%

\noindent \underline{Case(d) $k_{1R}<0$,$k_{2R}<0$}:
\begin{eqnarray}
\left(
\begin{array}{c}
A_1^{1-}\\
A_2^{1-} 
\end{array}
\right)
&=&
\left(
\begin{array}{c}
e^{\delta_2} \\
e^{\delta_2'}
\end{array}
\right)e^{-(R_2+R_3)/2},\\
\left(
\begin{array}{c}
A_1^{1+}\\
A_2^{1+} 
\end{array}
\right)
&=&
\left(
\begin{array}{c}
\alpha_1^{(1)} \\
\alpha_1^{(2)}
\end{array}
\right)e^{-R_1/2},\\
\left(
\begin{array}{c}
A_1^{2-}\\
A_2^{2-} 
\end{array}
\right)
&=&
\left(
\begin{array}{c}
\alpha_2^{(1)} \\
\alpha_2^{(2)}
\end{array}
\right)e^{-R_2/2},\\
\left(
\begin{array}{c}
A_1^{2+}\\
A_2^{2+} 
\end{array}
\right)
&=&
\left(
\begin{array}{c}
e^{\delta_1} \\
e^{\delta_1'}
\end{array}
\right)e^{-(R_1+R_3)/2},
\end{eqnarray}
$\hat{\phi}^{1-}=\frac{R_3-R_2}{2}$,$\hat{\phi}^{1+}=
\frac{R_1}{2}$,$\hat{\phi}^{2-}= \frac{R_2}{2}$
and $\hat{\phi}^{2+}
=\frac{R_3-R_1}{2}$.\\

%%%%%%%%%%%%%%%
\noindent In equations(6-21) the superscripts denote the solitons and the subscripts 
denote the modes. It is clear from the above expressions that in all 
the cases there exists a redistribution of intensity 
among the solitons. However it should be noticed that though there is an
intensity redistribution among the solitons in two modes, the total intensity of
individual soliton is conserved during collision process, that is,  
$|A_1^{j-}|^2+|A_2^{j-}|^2=|A_1^{j+}|^2+|A_2^{j+}|^2=\frac{1}{\mu}$, $j=1,2$,
which is of course a consequence of the integrability of the Manakov model. 
For example, the amplitude change in the two modes of soliton 1 
after interaction can be expressed by the following transformation, 
\begin{eqnarray}
A_1^{1+} &=& \Gamma C_{11}A_1^{1-} + \Gamma C_{12}A_2^{1-},\nonumber\\
A_2^{1+} &=& \Gamma C_{21}A_1^{1-} + \Gamma C_{22}A_2^{1-}.
\end{eqnarray}
Here $\Gamma$ $=$ $\Gamma$($A_1^{1-},A_2^{1-},A_1^{2-},A_2^{2-}$)
$\equiv$ $\left(\frac{a_2}{a_2^*}\right)$ $[1/((\alpha_1^{(1)}\alpha_2^{(1)*}+
\alpha_1^{(2)}\alpha_2^{(2)*})(\alpha_2^{(1)}\alpha_2^{(1)*}+
\alpha_2^{(2)}\alpha_2^{(2)*}))]$
$\left[\frac{1}{|\kappa_{12}|^2}-\frac{1}{\kappa_{11}\kappa_{22}}\right]^{-1/2}$, 
in which $a_2=(k_2+k_1^*)[(k_1-k_2)(\alpha_1^{(1)}\alpha_2^{(1)*}+
\alpha_1^{(2)}\alpha_2^{(2)*})]^{1/2}$. The forms of $C_{ij}$'s, $i,j=1,2$ read
as 
$C_{11} =$
$\alpha_2^{(1)}\alpha_2^{(1)*}(k_1-k_2)+\alpha_2^{(2)}\alpha_2^{(2)*}(k_1+k_2^*),
$ $C_{12} = -\alpha_2^{(1)}\alpha_2^{(2)*}(k_2+k_2^*)$, 
$C_{21} = -\alpha_2^{(2)}\alpha_2^{(1)*}(k_2+k_2^*)$,
$C_{22} =$
$\alpha_2^{(1)}\alpha_2^{(1)*}(k_1+k_2^*)+\alpha_2^{(2)}\alpha_2^{(2)*}(k_1-k_2)
$. Note that $C_{ij}$'s are independent of $\alpha_1^{(j)}$'s and so of
$A_1^{1-}$ and $A_2^{1-}$. 
Similar relations for the soliton 2 hold good for $A_1^{2+}$ and $A_2^{2+}$ 
also. 
Then the ratios of the $A_i^{j}$'s, $i,j=1,2,$ can be connected through
 linear fractional 
transformations(LFTs). For example, for soliton 1,
\begin{equation}
\rho_1^+=\frac{A_1^{1+}}{A_2^{1+}}=\frac{C_{11}\rho_1^-+C_{12}}
{C_{21}\rho_1^-+C_{22}},
\end{equation}
where $\rho_1$$=$$\frac{A_1^{1-}}{A_2^{1-}}$, ensuring that for every transformation there exists an inverse transformation. 
This idea has been profitably used in constructing logic gates[5,6]. In fact the
LFT (23) is identical to the LFT given by equation(9) in ref.[5] under the
change of notation $\rho_1^-$$\rightarrow$$\rho_1$,  
$\rho_1^+$$\rightarrow$$\rho_R$ with $C_{ij}$'s identified as the expressions
given therein.

\noindent Further, it is observed that the absolute value of the 
phase shift of the two solitons in all the above four cases is same  
and is given by 
\begin{eqnarray}
|\Phi| &=& \frac{|R_3-R_1-R_2|}{2},\nonumber\\
&=&\frac{1}{2}\mbox{log}\left[\frac{|k_1-k_2|^2(|\kappa_{11}
\kappa_{22}-\kappa_{12}\kappa_{21}|)}
{|k_1+k_2^*|^2\kappa_{11}\kappa_{22}}\right],\nonumber\\
%&=&\frac{1}{2}\mbox{log}\left[
% \frac{|k_1-k_2|^2(\bar{\kappa}_{11}\bar{\kappa}_{22}|k_1+k_2^*|^2-|\bar{\kappa}_{12}|^2
%(k_1+k_1^*)(k_2+k_2^*))}{|k_1+k_2|^4\bar{\kappa}_{11}\bar{\kappa}_{22}
%}\right], \nonumber\\
%%%%%%%%%%%%%
&=&\frac{1}{2}\mbox{log}\left[\frac{|k_1-k_2|^2}{|k_1+k_2^*|^4}\right]
\nonumber\\
&+&\frac{1}{2}\mbox{log}\left[
\frac{\bar{\kappa}_{11}\bar{\kappa}_{22}|k_1+k_2^*|^2-
|\bar{\kappa}_{12}|^2
(k_1+k_1^*)(k_2+k_2^*)}
{\bar{\kappa}_{11}\bar{\kappa}_{22}}
\right], \nonumber\\
%%%%%%%%%%%%%
%&=&\frac{1}{2}\mbox{log}\left[ \frac{\bar{\kappa}_{11}\bar{\kappa}_{22}|k_1+k_2^*|^2}
%{\bar{\kappa}_{11}\bar{\kappa}_{22}|k_1+k_2^*|^2-|\bar{\kappa}_{12}|^2
%(k_1+k_1^*)(k_2+k_2^*)}\right], \nonumber\\
\end{eqnarray}
where $\bar{\kappa}_{ij} = \mu(\alpha_i^{(1)}\alpha_j^{(1)*}+
\alpha_i^{(2)}\alpha_j^{(2)*}),\;i,j=1,2.$ and the absolute value of 
the change in relative separation 
distance  
$t_{12}^{\pm}$ (position of $S_2$ (at $z \rightarrow \pm \infty)$
- position 
of $S_1$ (at $z \rightarrow \pm \infty)$) is given by
\begin{equation}
|\Delta t_{12}|=|t_{12}^- - t_{12}^+|=\left|\frac{(k_{1R}+k_{2R})}
{2k_{1R}k_{2R}}\right|
|\Phi|.
\end{equation}
It is interesting to note that in the collision process the phase shift 
is not only dependent on the $k_j$'s, $j=1,2$ but also on the 
complex parameters $\alpha_i^{(j)}$'s, $i,j=1,2$, and so on $A_i^j$'s. These two 
properties, that is, dependence of the change in the intensity profiles 
of the solitons in the two modes and of the phase shift of them during 
collision on the parameters $\alpha_i^{(j)}$'s  
make the collision properties novel, not seen in general other standard $(1+1)$ 
dimensional 
soliton systems. 

\noindent Now looking at the dependence of the phase shift on 
$\alpha_i^{(j)}$'s, we can consider two special cases.

\noindent Case(a):
\underline{$\alpha_1^{(1)}:\alpha_2^{(1)}=
\alpha_1^{(2)}:\alpha_2^{(2)}$}.

\noindent In this case(corresponding to parallel modes) the collision 
correspond to pure elastic collision ($|A_i^{j+}|=|A_i^{j-}|$, $i,j=1,2$) and the 
phase shift is given by 
\begin{equation}
\left|\Phi\right|=\left|\mbox{log}
\left[\frac{|k_1-k_2|^2}{|k_1+k_2^*|^2}\right]\right|.
\end{equation}
Case(b):
\underline{$\alpha_1^{(1)}:\alpha_2^{(1)}=$$\infty$, 
$\alpha_1^{(2)}:\alpha_2^{(2)}=0$}.\\
\noindent This case corresponds to two orthogonal modes. Here the phase shift is 
given by
\begin{equation}
\left|\Phi\right|=\left|\mbox{log}\left[\frac{|k_1-k_2|}{|k_1+k_2^*|}
\right]\right|.
\end{equation}
These two examples show that the phase shifts and hence the relative 
separation distances (see eq.(25)) vary as the complex parameters $\alpha_i^{(j)}$'s 
change and this variation will be reflected in the shape of the profiles 
of the two interacting solitons.
%%%%%%%%%%%%%%%%%%%%%%%
\section{Stationary solitons and relative separation distances}
In order to realize the effect of phase shift on the shape of the solitons 
we consider the simple case of stationary limit of the two-soliton solution 
(5). Let us consider the situation in which the velocities are zero, that is 
$k_{jI}=0$, $j=1,2$. Further we choose 
$\alpha_2^{(1)}=\alpha_1^{(2)}=0,$ $\alpha_1^{(1)} = e^{\eta_{10}}, $
$\alpha_2^{(2)} = -e^{\eta_{20}}$ and $k_{nI}=0$, where $\eta_{i0},i=1,2$ 
are real parameters (corresponding to orthogonal modes). 
In this limit, the two-soliton solution(5) reduces to 
\begin{eqnarray}
q_1 &=& 2k_{1R}\sqrt{\frac{k_{1R}+k_{2R}}{k_{1R}-k_{2R}}}\mbox{cosh}
(k_{2R}\bar{t_2}) 
e^{ik_{1R}^2z}/D_1,\\
q_2 &=& 2k_{2R}\sqrt{\frac{k_{1R}+k_{2R}}{k_{1R}-k_{2R}}}\mbox{sinh}
(k_{1R}\bar{t_1}) 
e^{ik_{2R}^2z}/D_1\\
D_1&=&\sqrt{\mu}\mbox{cosh}(k_{1R}\bar{t_1}+k_{2R}\bar{t_2})
+\sqrt{\mu}\left(\frac{k_{1R}+k_{2R}}{k_{1R}-k_{2R}}\right)\nonumber\\
&&\mbox{cosh}(k_{1R}\bar{t_1}-k_{2R}\bar{t_2}) ,\\
\bar{t_1}&=&t-t_1=t+\frac{\eta_{10}}{k_{1R}}+\frac{1}{2k_{1R}}
\mbox{log}\left[\frac{\mu(k_{1R}-k_{2R})}{4k_{1R}^2(k_{1R}+k_{2R})}
\right],\nonumber\\
\\
\bar{t_2}&=&t-t_2=t+\frac{\eta_{20}}{k_{2R}}+\frac{1}{2k_{2R}}
\mbox{log}\left[\frac{\mu(k_{1R}-k_{2R})}{4k_{2R}^2(k_{1R}+k_{2R})}
\right],\nonumber
\\
\end{eqnarray}
so that each soliton is in one particular mode.

\noindent It is also of interest to note that the stationary limit of the two-soliton 
solution obtained above is also the so-called partially coherent stationary 
soliton studied in the literature intensively in recent times [8,10] in 
connection with the existence of spatial solitons in photorefractive materials. 
In fact eqs.(28-32) are nothing but the stationary 2-PCS solution 
obtained in eqs.(13-15) 
in ref.[10].
Now we can identify the relative separation distance 
between the solitons as
 \begin{eqnarray}
 t_{12} = t_2-t_1 &=& \frac{\eta_{10}}{k_{1R}}-
\frac{\eta_{20}}{k_{2R}}+\frac{1}{2k_{1R}}
\mbox{log}\left[\frac{\mu(k_{1R}-k_{2R})}{4k_{1R}^2(k_{1R}+k_{2R})}\right]\nonumber\\
&&-\frac{1}{2k_{2R}}
\mbox{log}\left[\frac{\mu(k_{1R}-k_{2R})}{4k_{2R}^2(k_{1R}+k_{2R})}
\right].
\end{eqnarray}
It can be very easily seen that the shape of the stationary soliton 
depends very much on the relative separation distance $t_{12}$.
To illustrate this in fig.(1) we have plotted (i)symmetric case $t_{12}=0$ 
and (ii) asymmetric case $t_{12}\neq0$.

\noindent One can proceed to consider the propagation of the above stationary 
soliton solution and check their shape changing properties under 
collision. Choosing the parameters $\alpha_1^{(2)}$ and 
$\alpha_2^{(1)}$ as functions 
of velocities ($k_{jI}$) such that they vanish when $k_{jI}=0$, $j=1,2$,
 the nature of soliton 
collisions is shown in figs.2 for the parameters 
$\alpha_1^{(1)}=1.0$,$\;$$\alpha_1^{(2)}=1.0$,$\;$
$\alpha_2^{(1)}=\frac{22+80i}{89}$,$\;$$\alpha_2^{(2)}=-2.0$,$\;$
$k_1=1.0+i$ and $k_2=2.0-i$.

\section{Conclusion}
The soliton interactions in CNLS equations possess very rich structure. 
In this paper, we have discussed the two-soliton interaction properties 
of 2-CNLS equations with special emphasis on the nature of phase-shift 
encountered by 
solitons under collision and its dependence on the amplitudes of the 
modes. We have also pointed out that the much discussed stationary PCS
solitons correspond to stationary limit of appropriate soliton solutions 
with 2-PCS as an example here. 
Because of the complex nature of soliton interaction, 
multisoliton solution in multicomponent systems possess highly 
nontrivial structures. These properties will be presented seperately [12]. 
Such studies are expected to have very important application in optical 
communications, optical devices and optical computing.

\noindent {\bf Acknowledgemet}

\noindent The work of M. L. and T. K. forms a part of a Department of Science and 
Technology, Government of India, funded research project.

\begin{figure*}[h]
\resizebox{0.75\textwidth}{!}{%
  \includegraphics{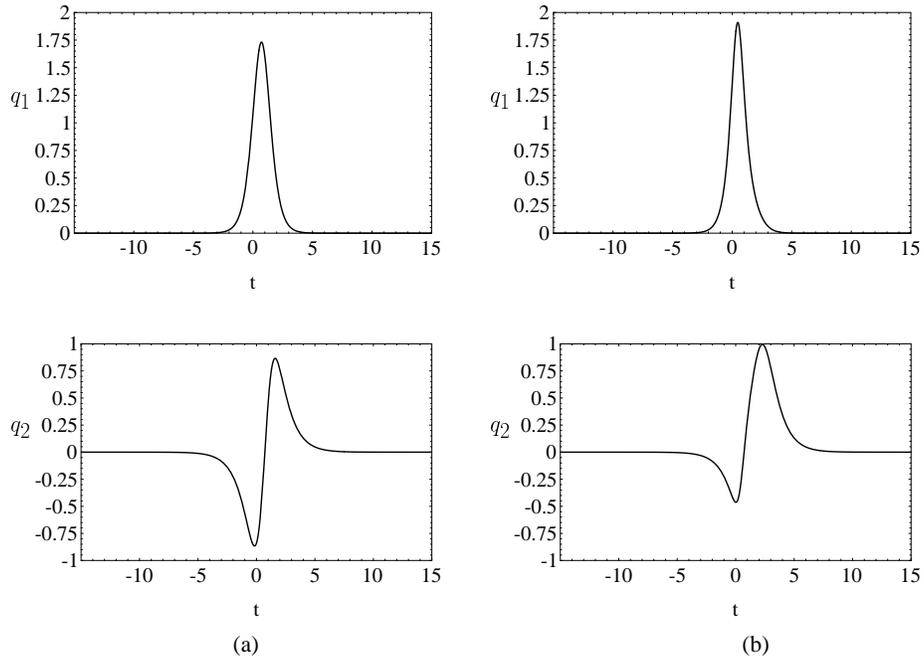}
	}
\caption{Typical stationary form of the 2-soliton solution (PCS)
for the 2-CNLS system for z=0,
see eqs.(28-32) :
(a) symmetric case ($t_{12}=0)$, (b) asymmetric case 
($t_{12}=1.0$).}
\label{fig:1}      
\end{figure*}
\begin{figure*}[h]
\resizebox{0.75\textwidth}{!}{%
  \includegraphics{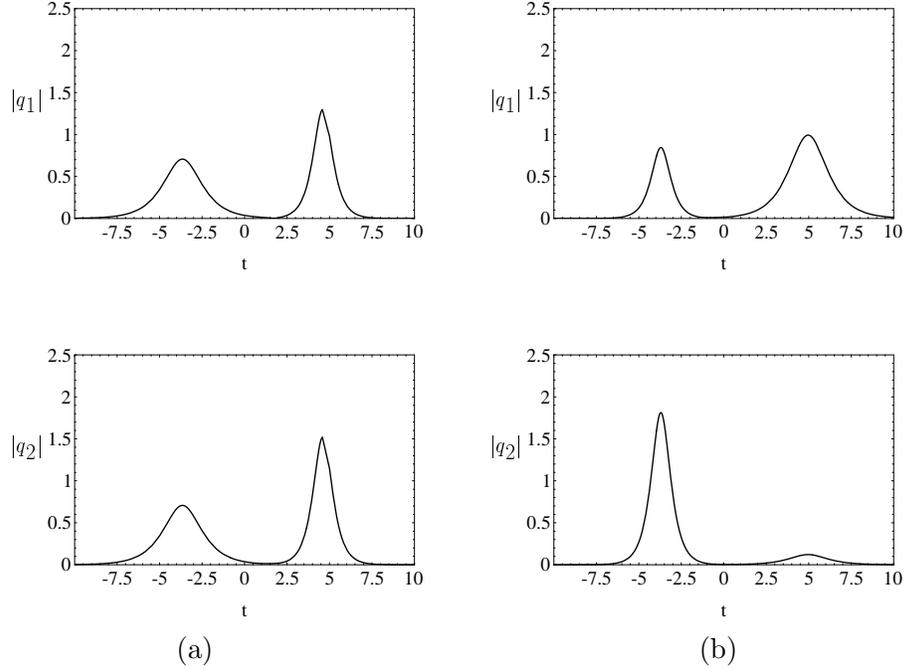}
	}
\hspace{-290pt}(a) \hspace{180pt}(b)
\caption{Asymptotic forms of two-soliton solution (whose stationary form is 
similar to fig.1) 
of the integrable 2-CNLS equations (a) at z=-2 and 
(b) at z=2.}
\label{fig:1}      
\end{figure*}


\begin{thebibliography}{99}
\bibitem{re1}
G. P. Agrawal, {\textit Nonlinear Fiber Optics-Second Edition} 
(Academic Press, New York, 1995).
\bibitem{re2}
See for example, several articles in the Focus Issue on ``Optical 
Solitons - Perspectives and Applications" in 
Chaos {\bf 10}, No.3 (2000).
\bibitem{re3p}
S. V. Manakov, Zh. Eksp. Teor. Fiz. {\bf 65}, (1973) 505 [Sov. Phys.
JETP {\textbf 38}, (1974) 248 ].
\bibitem{re3}
R. Radhakrishnan, M. Lakshmanan, and J. Hietarinta, Phys. Rev.  E
{\textbf56}, (1997) 2213.
\bibitem{re4}
M. H. Jakubowski, K. Steiglitz and R. Squier,
Phys. Rev. E {\textbf 58}, (1998)  6752.
\bibitem{re5}
K. Steiglitz, Phys. Rev. E. {\textbf 63}, (2000) 016608 .
\bibitem{re7}
J. Yang, Phys. Rev. E {\textbf 59}, (1999) 2393; 
Phys. Rev. E {\textbf 64}, (2001) 026607.
\bibitem{6}
M. Mitchell and M. Segev, 
Nature, {\textbf 387}, (1997)  880.
\bibitem{re7}
T. Kanna and M. Lakshmanan, Phys. Rev. Lett, {\textbf 86}, (2001) 5043 .
\bibitem{re8}
A. Ankiewicz, W. Krolikowski, and N. N. Akhmediev, Phys. Rev. E 
{\textbf 59},  (1999) 6079.
\bibitem{re9}
M. Lakshmanan, T. Kanna and R. Radhakrishnan, Rep. Math. Phys. {\textbf 46},
(2000) 143 .
\bibitem{re10}
T. Kanna and M. Lakshmanan, to be published.
\end{thebibliography}
\end{document}